\documentclass[fleqn,twoside]{article}
\usepackage[headings]{espcrc2}

\usepackage{graphicx}
\usepackage[figuresright]{rotating}
\usepackage{pstricks}
\usepackage{amsmath}

\newcommand{\be}{\begin{equation}}
\newcommand{\ee}{\end{equation}}
\newcommand{\bea}{\begin{eqnarray}}
\newcommand{\eea}{\end{eqnarray}}

\newcommand{\lesssim}{ {\
\lower-1.2pt\vbox{\hbox{\rlap{$<$}\lower5pt\vbox{\hbox{$\sim$}}}}\ } 
}
\newcommand{\gtrsim}{ {\
\lower-1.2pt\vbox{\hbox{\rlap{$>$}\lower5pt\vbox{\hbox{$\sim$}}}}\ } 
}

\title{Asymptotic expansions of Feynman diagrams and the Mellin-Barnes representation}

\author{ Samuel Friot \address[CPT]{Centre  de Physique Th{\'e}orique~CNRS-Luminy, Case 907 
F-13288 Marseille Cedex 9, France \\ CNRS UMR 6207} \address[LM]{Laboratoire de Math\'ematiques Appliqu\'ees du Havre, Universit\'e du Havre UFR des Sciences et techniques - 25, rue Philippe Lebon  BP 540, 76058 Le Havre, France} \thanks{Talk given at the $12^{th}$ 
High-Energy Physics International Conference on Quantum Chromodynamics, 4-8 July (2005), Montpellier 
(France).}, David Greynat$^{\,\, \rm a}$}

\runtitle{Asymptotic expansions of Feynman diagrams and the Mellin-Barnes representation.}
\runauthor{S. Friot, D. Greynat}

\begin{document}

\begin{abstract}
In this talk, we describe part of our recent work \cite{FGdeR05} (see also \cite{F05,G05}) that gives new results in the context of asymptotic expansions of Feynman diagrams using the Mellin-Barnes representation.
\vspace{1pc}
\end{abstract}

\maketitle
\section{Introduction}\label{int}
In the Standard Model quantum field theories, where calculations of Feynman diagrams are often reduced to the evaluation of complicated integrals involving several scale parameters, exact results are difficult to obtain. It is therefore legitimate to search to compute asymptotic expansions of these integrals in the various relevant limits of the parameters (or ratios of them). A powerful mathematical strategy to compute asymptotic expansions of integrals depending on parameters (in the limits of small and/or large values of the parameters) is based on the Mellin-Barnes representation \cite{W90,KP01}. The use of Mellin-Barnes representation in quantum field theory goes back in ref.\cite{BW63} and many applications have been made since then. In ref.\cite{FGdeR05} we have proposed a new way for performing calculations within this strategy, using a theorem and new mathematical objects recently introduced in algorithmics \cite{FGD95}. 
In this talk, we shall briefly describe these mathematical tools and show the power of our approach on a few diagrams of interest for the investigation of the anomalous magnetic moment of the muon. As a pedagogical example, a very elementary calculation will be exposed in a detailed way.

\vspace{-0.3cm}

\section{General lines of the strategy and some mathematical tools}

The underlying mathematical procedure of our method can be nicely summarized in only one sentence \cite{FGD95}: \textit{"For estimating asymptotically a function $f(x)$} [in the limit of small or large $x$]\textit{, determine its Mellin transform $\mathcal{M}[f](s)$ and translate back its singularities into asymptotic terms in the expansion of $f(x)$"}.

Therefore, in the case of a Feynman diagram, we see that the problem is equivalent to get the Mellin-Barnes representation of its corresponding Feynman integral in a first step (since the Mellin-Barnes representation is the other denomination of the inverse Mellin transform) and once done, in a second step, to care for the singular structure of its explicitly obtained Mellin transform. 

The first step can be obtained in an easy manner: first use the Feynman parameterisation, then compute the momenta integrals to get purely parametric integrals and finally introduce an appropriate Mellin-Barnes representation of part of the integrand (or several if one has several convenient parameters). Then, commute integrals over Feynman parameters and the Mellin-Barnes representation(s) and compute the parametric integrals. In this way, the result is a Mellin-Barnes representation of the diagram where the Mellin transform of the diagram, directly readable in the integrand, is in general a product of Euler's Gamma fonctions. In some of the examples that we have considered, the cosecant function also appeared. In these cases, it can be more convenient to keep this function explicitly rather than to express it in terms of Gamma functions, since it can supply useful simplifications of the calculations.

To deal with the second step of our problem, we need to define two important mathematical objects (the \textit{fundamental strip} and the \textit{singular expansion} of the Mellin transform of our diagram) and to give a theorem (the \textit{converse mapping theorem}). These tools have not been used previously in quantum field theory calculations but their efficiency in this context was clearly shown in our work \cite{FGdeR05,F05,G05}. They are going to provide the dictionary between the singular structure of the Mellin transform of the diagram and its asymptotic expansions.

\vspace{0.4cm}

The \textit{fundamental strip} $<\alpha,\beta>$ of the Mellin transform $\mathcal{M}[f](s)$ of a locally integrable function $f(x)$ (where $x$ is a positive real number) is the largest open strip of the complex $s$-plane in which the integral 
\begin{equation}\nonumber
\mathcal{M}[f](s)=\int_0^\infty dx x^{s-1}f(x)
\end{equation}
is convergent \cite{FGD95}.

This fundamental strip can be easily found from the following conditions:

If $f(x)\underset{x\rightarrow0^+}=\mathcal{O}(x^{-\alpha})$ and $f(x)\underset{x\rightarrow+\infty}=\mathcal{O}(x^{-\beta})$ with $\alpha>\beta$ then $\mathcal{M}[f](s)$ converges in $<\alpha,\beta>$ (an alternative notation for the fundamental strip is $\Re(s)\in]\alpha,\beta[$).

\vspace{0.4cm}

The \textit{singular expansion} of the Mellin transform in a given region of the complex $s$-plane,  outside of the fundamental strip and where it admits a meromorphic continuation, is a formal sum collecting all singular elements of the Mellin transform in this region, a singular element (at a pole $p$) being simply the truncated Laurent expansion of the Mellin transform, at order $\mathcal{O}\left((s-p)^0\right)$ or less (depending on the Laurent expansion), around this pole \cite{FGD95}. We note this formal series 
\begin{equation}\nonumber
\mathcal{M}[f](s)\asymp\sum_{p,n}a_{p,n}\frac{1}{(s-p)^n}\,.
\end{equation}

Finally the \textit{converse mapping theorem} is the following \cite{FGD95}.

Let $f(x)$ be continuous in $]0,+\infty[$ with Mellin transform $\mathcal{M}[f](s)$ having a non empty fundamental strip $<\alpha,\beta>$. 

$(i)$ If, for some $\gamma<\alpha$, $\mathcal{M}[f](s)$ admits a meromorphic continuation to the strip $<\gamma,\alpha>$, if there exists $\eta\in]\alpha,\beta[$ such that, in $\gamma\leq\Re(s)\leq\eta$, $\mathcal{M}[f](s)\underset{|s|\rightarrow\infty}=\mathcal{O}(|s|^{-r})$ with $r>1$, and if $\mathcal{M}[f](s)$ admits the singular expansion 
\begin{equation}\nonumber
\mathcal{M}[f](s)\asymp\sum_{p,n}a_{p,n}\frac{1}{(s-p)^n}
\end{equation}
in $<\gamma,\alpha>$, then an asymptotic expansion of $f(x)$ at $0$ is
\begin{equation}\nonumber
f(x)=\sum_{p,n}a_{p,n}\left[\frac{(-1)^{n-1}}{(n-1)!}x^{-p}\ln^n(x)\right]+\mathcal{O}(x^{-\gamma}).
\end{equation}

$(ii)$ Similarly if, for some $\gamma>\beta$, $\mathcal{M}[f](s)$ admits a meromorphic continuation to the strip $<\beta,\gamma>$, if there exists $\eta\in]\alpha,\beta[$ such that, in $\eta\leq\Re(s)\leq\gamma$, $\mathcal{M}[f](s)\underset{|s|\rightarrow\infty}=\mathcal{O}(|s|^{-r})$ with $r>1$, and if $\mathcal{M}[f](s)$ admits the singular expansion 
\begin{equation}\nonumber
\mathcal{M}[f](s)\asymp\sum_{p',n'}b_{p',n'}\frac{1}{(s-p')^{n'}}
\end{equation}
in $<\beta,\gamma>$, then an asymptotic expansion of $f(x)$ at $+\infty$ is
\begin{equation}\nonumber
f(x)=\sum_{p',n'}b_{p',n'}\left[\frac{(-1)^{n'}}{(n'-1)!}x^{-p'}\ln^{n'}(x)\right]+\mathcal{O}(x^{-\gamma}).
\end{equation}

In summary, each term of the left (resp. right)\footnote{By left (resp. right) we mean left (resp. right) compared to the fundamental strip.} singular expansion of the Mellin transform of the function gives a corresponding term of the asymptotic expansion of the function in the zero (resp. infinite) limit of $x$ by simple reading.

\vspace{0.4cm}

Some comments are in order. 

First, the fundamental strip of the Mellin transform is very important, because it separates two distinct areas of the complex $s$-plane : a left and a right one. In the \textit{left} region, the poles of the Mellin transform are linked, by the converse mapping theorem, to the asymptotic expansion of the original function $f(x)$ in the \textit{zero} limit of $x$. But poles in the \textit{right} region are linked to the asymptotic expansion of $f(x)$ in the \textit{infinite} limit of $x$. It is therefore crucial to identify the fundamental strip of the Mellin transform in order to know which poles are associated to which asymptotic expansion. This is in general easy in the case of Feynman diagrams. As described above, when writing the Mellin-Barnes representation of a diagram (after that the corresponding Feynman integral has already been transformed into purely parametric integrals), one uses the Mellin-Barnes representation of part of the parametric integrand, so that one knows the fundamental strip for this expression (using the rules explained above). Notice however that the evaluation of the parametric integrals, once their commutation with the Mellin-Barnes representation realized, can reduce the size of the fundamental strip. Therefore, the fundamental strip of the Mellin transform of a diagram is obtained only after evaluating the parametric integrals. 

Secondly, it is not necessary to compute the whole singular expansion(s) if one is only interested in the first few terms, or in one particular term, of the asymptotic expansion(s). In these cases, one only has to consider the closest poles to the fundamental strip, or the pole corresponding to the particular term, in the relevant half complex plane.

Finally, in many cases (in particular in all cases that we have considered so far), Mellin transforms admit a meromorphic continuation in the complete left and/or right half complex plane, therefore complete asymptotic expansion(s) for the original function can be found. Such expansions can be either convergent or divergent and, if convergent, they may \textit{in some cases} represent the function exactly \cite{FGD95}.

\vspace{-0.3cm}
\section{Applications: the anomalous magnetic moment of the muon}

The anomalous moment of the muon is a well-known precision observable for testing the Standard Model (for a recent review, see \cite{Passera}).

One can easily use our method to compute the asymptotic expansions of the contributions to the anomaly of the diagrams of figure below (where $\ell$ and $\ell'$ are either electrons or taus with $\ell\neq \ell'$, and $H$ denotes a hadronic polarisation\footnote{Higher order hadronic polarisation contributions to the muon anomaly have first been studied in \cite{Calmet:1976kd}.} $\Pi_H$). We have therefore been able, among others, to directly check in \cite{FGdeR05,F05,G05} the results of \cite{Czar} (diagram 2 of our figure) and eq. (8) of \cite{Krause} corresponding to diagrams 4 and 5 (notice that eq. (8) of \cite{Krause} is \textit{not} valid for a tau in the electromagnetic polarisation). To be more precise, diagram 1 can be computed in the $\frac{m_\ell^2}{m_\mu^2}\rightarrow +\infty$ or $\frac{m_\ell^{2}}{m_\mu^2}\rightarrow 0$ limits and diagram 2 in both limits simultaneously. In the case of the other diagrams involving a hadronic polarisation, perturbative and non-perturbative contributions are involved and it is therefore better to separate them with the use of a dispersion relation written for the hadronic polarisation. One can then apply the asymptotic expansion procedure on the associated kernel function in the limit $\frac{t}{m_\mu^2}\rightarrow+\infty$ where $t$ is the dispersive integration variable.
\begin{center}
\includegraphics[width=0.46\textwidth]{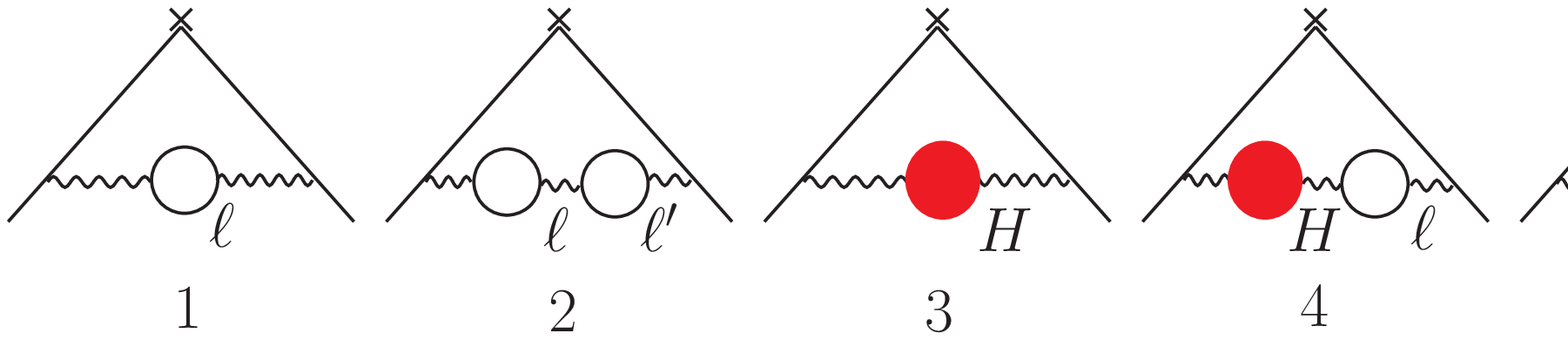}
\end{center}
As a pedagogical example, let us consider one of the simplest of these diagrams in a detailed manner, namely diagram number 3.

On can easily show that, once the dispersion relation has been worked out for $\Pi_H$, the contribution of this diagram to the anomaly is
$F_{2}\left( 0\right)=\frac{\alpha}{\pi}\int_{4m_\pi^2}^{+\infty}dt\Im(\Pi_H(t))K_2(\frac{t}{m_\mu^2})$, where
\begin{equation}\nonumber
K_2\left(\frac{t}{m_\mu^2}\right) =\int _{0}^{1}dx\frac{1-x}{1+\frac{(1-x)}{x^2}\, \frac{t}{m_\mu^2}}\,
\end{equation}
is a well-known kernel function. We are interested in the asymptotic expansion of $K_2\left(\frac{t}{m_\mu^2}\right)$ when $\frac{t}{m_\mu^2}\rightarrow+\infty$. Although it is trivial to compute it from the exact result of the kernel function that is easy to obtain, let us show how to manage this evaluation within our strategy.

First, introduce the Mellin-Barnes representation
\begin{equation}\label{MB}
\frac{1}{1+X}=\frac{1}{2i\pi}\int_{c-i\infty}^{c+i\infty}ds\, X^{-s}\frac{\pi}{\sin(\pi s)}
\end{equation}
of fundamental strip $<0,1>\ni c$ and where $X=\frac{(1-x)}{x^2}\, \frac{t}{m_\mu^2}$.
One then gets
\begin{multline}\hspace{-0.5cm}
\nonumber
K_2\left(\frac{t}{m_\mu^2}\right) =\frac{1}{2i\pi}\int_{c-i\infty}^{c+i\infty}ds\left(\frac{t}{m_\mu^2}\right)^{-s}\frac{\pi}{\sin(\pi s)}\\
\times\int _{0}^{1}dx\,x^{2s}(1-x)^{1-s}\,.
\end{multline}
This last equation involves the integral representation of the Beta function $B(2s+1,2-s)$ which is equal to $\frac{\Gamma(2s+1)\Gamma(2-s)}{\Gamma(s+3)}$ if $\Re(2s+1)>0$ and $\Re(2-s)>0$. These conditions being always satisfied, since $\Re(s)\in]0,1[$ (the fundamental strip of eq. (\ref{MB})), we finally find the Mellin-Barnes representation of the kernel function
\begin{multline}\hspace{-0.5cm}
\nonumber
K_2\left(\frac{t}{m_\mu^2}\right)=\frac{1}{2i\pi}\int_{c-i\infty}^{c+i\infty}ds\left(\frac{t}{m_\mu^2}\right)^{-s}\\
\times\frac{\pi}{\sin(\pi s)}\frac{\Gamma(2s+1)\Gamma(2-s)}{\Gamma(s+3)}\,.
\end{multline}
Notice that in this case, the fundamental strip has not been perturbed after the integration over the Feynman parameter. The Mellin transform of the kernel function is then
\begin{equation}
\label{mellin}
\mathcal{M}\left[K_2(\frac{t}{m_\mu^2})\right](s)=\frac{\pi}{\sin(\pi s)}\frac{\Gamma(2s+1)\Gamma(2-s)}{\Gamma(s+3)}\, 
\end{equation}
of fundamental strip $<0,1>$.

Since we want the asymptotic expansion of $K_2$ in the infinite limit of $\frac{t}{m_\mu^2}$, we need to compute the singular expansion of (\ref{mellin}) to the right of the fundamental strip.
Looking at (\ref{mellin}), we see that it can be easily obtained by calculating its truncated Laurent series around an arbitrary strictly positive natural integer $n$ (the only singular functions of (\ref{mellin}) in the right half complex plane being the cosecant and the $\Gamma(2-s)$ functions). One finds
\begin{multline}\hspace{-0.5cm}
\nonumber
\mathcal{M}\left[K_2\left(\frac{t}{m_\mu^2}\right)\right](s)\asymp-\frac{1}{3}\frac{1}{s-1}\\
+\sum_{n=2}^{\infty}\left\{\left[\frac{1}{s-n}\right]^2A_n+\frac{1}{s-n}\,B_n\right\}\,,
\end{multline}
with 
\begin{equation}
\hspace{-1cm}
\small{A_n=-\frac{1}{(n-2)!} \frac{\Gamma(2n+1)}{\Gamma(n+3)}} \nonumber\\ 
\end{equation}
\begin{equation}
\hspace{-1cm}
\small{B_n=-A_n\left[\psi(n-1)-2\psi(2n+1)+\psi(n+3)\right]\,,}\nonumber
\end{equation}
where $\psi$ is the Digamma function.

Then, applying the converse mapping theorem, we finally have the asymptotic expansion
\begin{equation}
\nonumber
K_2\left(\frac{t}{m_\mu^2}\right)\underset{\frac{t}{m_\mu^2}\rightarrow\infty}=\frac{1}{3}\frac{m_\mu^2}{t}
\end{equation}
\begin{equation}\nonumber\hspace{-0.2cm}
+\sum_{n=2}^{\infty}\left\{A_n\left(\frac{t}{m_\mu^2}\right)^{-n}\ln\frac{t}{m_\mu^2}+(-1)B_n\left(\frac{t}{m_\mu^2}\right)^{-n}\right\}\,.
\end{equation}

Our strategy is, as we said, also useful for the other diagrams of the above figure. For diagrams 2, 4 and 5, calculations are slightly more complicated but remain still easy to perform \cite{FGdeR05,F05,G05}.

\end{document}